# Noble metals on anodic $TiO_2$ nanotubes mouths: Thermal dewetting of minimal Pt co-catalyst loading leads to significantly enhanced photocatalytic $H_2$ generation

*Nhat Truong Nguyen, Marco Altomare, JeongEun Yoo, Nicola Taccardi, and Patrik Schmuki\**

N. T. Nguyen, Dr. M. Altomare, J. E. Yoo, Prof. Dr. P. Schmuki
Department of Materials Science and Engineering WW4-LKO, University of Erlangen-Nuremberg, Martensstrasse 7, Erlangen, D-91058, Germany
E-mail: schmuki@ww.uni-erlangen.de

Dr. N. Taccardi
Lehrstuhl für Chemische Reaktionstechnik, University of Erlangen-Nuremberg, Egerlandstrasse 3, Erlangen, D-91058, Germany

Prof. Dr. P. Schmuki
Department of Chemistry, King Abdulaziz University, Jeddah, Saudi Arabia

Ever since the groundbreaking study of Fujishima and Honda,[1] $H_2$ production by photocatalytic splitting of $H_2O$ on semiconductor materials has been intensively investigated and envisaged as promising pathway for the production of environmentally friendly energy sources.[2–6] The photocatalytic process is essentially based on the absorption of light by a suitable semiconductor material which leads in turn to photo-promotion of electrons towards the conduction band, leaving behind holes in the valence band. After electron-hole pair separation, the charge carriers can react at the semiconductor surface with suitable redox species in the environment. Concerning hydrogen production through water splitting, essential is that the excited electron (that is, the energetic position of the conduction band) is able to reduce water to $H_2$.

In spite of numerous types of explored semiconductors, titanium dioxide ($TiO_2$) remains the most investigated material for photocatalysis and ("open-circuit") $H_2$ production. $TiO_2$ has in fact a suitable conduction band energy, is cheap, non-toxic and, most importantly, is stable against corrosion and photocorrosion.[7,8]

In general, to maximize the efficiency of a (photo)catalytic process, nanostructured (photo)catalysts are employed, that are commonly based on nanoparticles or nanoparticle aggregates.[9,10] Not only does this provide a large surface area (this typically leads to large number of reaction sites) and thus to a higher specific yield, but also for photocatalytic processes the recombination of photo-induced charge carriers (detrimental to the overall efficiency of the process) is drastically reduced, as electrons and holes only have to diffuse (migrate) over distances in the nm-scale to reach the solid-liquid interface and react with the capturing environment.

In the recent years, instead of particles, one-dimensional morphologies such as nanotubes, nanowires and nanorods, have attracted particular interest in photocatalytic applications due to beneficial effects given by directional charge transport and orthogonal electron-hole separation.[8,9,11] Particularly, anodic self-organized TiO$_2$ nanotubes (NTs) have been intensively studied for a range of photo(electro)chemical applications, since entirely aligned arrays of NTs with controllable morphology can be grown by a simple electrochemical process. Additionally, these anodic NTs can be converted into the photo-electrochemically most active TiO$_2$ polymorph, i.e., anatase, by a simple annealing step.

Important however is to note that the H$_2$ evolution reaction, which implies the reduction of water by photo-generated electrons in the TiO$_2$ conduction band, is kinetically hampered[12,13] and typically can efficiently proceed only if a co-catalyst such as Pt, Au, Pd is deposited on the semiconductor.[14–17] These co-catalysts, the most efficient of which is Pt, enable efficient electron transfer at the interface and enhance the hydrogen recombination reaction (2H$^0$ → H$_2$).[18–20] Various works reported on the use of differently prepared Pt particles and investigate their *optimum* loading, particle size and distribution on TiO$_2$ powder, this to optimize the photocatalytic efficiency while minimizing the use of high-cost noble metal.[20] Also in the case of TiO$_2$ NT layers, different Pt decoration methods have been described,[21–23] mostly leading to a decoration of the full tube length with nanoparticles – one of the most frequently used and most efficient technique is photo-deposition.[24,25]

In the present work, we take a different approach and decorate by sputtering only the outermost surface of anodic TiO$_2$ NTs with very thin Pt films. Then we heat-treat these Pt/TiO$_2$ NT layers in optimized conditions to induce a controlled thermal dewetting of the Pt film. We show that optimally dewetted Pt films of a thickness of 1 nm and located only at the mouth of TiO$_2$ NTs lead to drastically enhanced photocatalytic H$_2$ generation, this compared to both non-dewetted tubes and to tubes that carry along their full length a classic Pt nanoparticle decoration - thus we achieve a higher efficiency with a much smaller Pt loading.

To fabricate the TiO$_2$ NT arrays we anodized titanium foils in an ethylene glycol based electrolyte containing 0.15 M NH$_4$F and 3 wt% H$_2$O, at 60 V for 15 min (see experimental section for more details). The NT layers were then annealed at 450 °C in air to convert the tubes to anatase. These anodic layers after formation are ~ 6.5 μm thick (**Figure 1**a), and consist of ordered arrays of NTs with individual diameter of ~ 120 nm (Figure 1b). At their top, these tubes carry a thin (*ca.* 80 nm thick) initiation layer (see the caption between the inset of Figure 1a and b). This initiation layer, as discussed below, can be regarded as a collar at the mouth of the tube and plays a crucial role in the localization of the Pt deposits just onto the tube mouth.

After tube formation, thin Pt layers of different nominal thickness (1, 2, 5 and 10 nm) were sputter-deposited onto the tube tops. Then the decorated tubes were treated at 550 °C in Ar, which leads to dewetting of the Pt layer into Pt particles (Figure 1c). The example in Figure 1 shows dewetting of a nominally 1 nm thick Pt film. The formed particles in this case have an individual size of approximately 4 nm and are uniformly distributed over the top surface. A statistical evaluation of the Pt particles yields a quite narrow size distribution (Figure 1d). In line with the transmission electron microscope (TEM) image in Figure 1e, a cross section investigation shows that the penetration of the Pt decoration reaches a depth of ~ 200 nm from the tube top (Figure S1d, Supporting Information). This low penetration can to a large extent be ascribed to the presence of the initiation layer (collar) at the tube top which prevents the Pt nanoparticles from being sputter-deposited deeply inside the tubes.

As anticipated, such dewetting experiments were carried out for various amounts of sputtered Pt (1 – 10 nm, see the Supporting Information). Well in line with theory on the dewetting mechanism,[26] the mean size of the Pt nanoparticles after thermal dewetting depends directly on the thickness of the initially deposited Pt layer (Figure S2 and 3, Supporting Information).

It should be pointed out that Ar annealing is essential for Pt dewetting to occur, while for instance, annealing in air atmosphere results in an unchanged morphology of the Pt film. One may assume that the thermal treatment in $O_2$-containing atmosphere could induce Pt passivation,[27,28] which hampers the occurrence of dewetting (i.e., limited surface-diffusion of Pt atoms in Pt oxide).[14]

High resolution transmission electron microscope (HR-TEM) analysis (Figure S4, Supporting Information) reveals that after the thermal treatment the NTs consist of an anatase – rutile mixed-phase. Particularly, anatase (101) and rutile (110) crystallographic planes, with a lattice spacing of 3.49 and 3.26 Å, respectively, could be observed.[29,30]

In order to assess the photocatalytic $H_2$ evolution performance of different thermally-dewetted structures, these were illuminated with UV light (60 mW cm$^{-2}$, 325 nm) in ethanol-water mixtures. Figure 1f and g show the $H_2$ evolution rates for Pt/$TiO_2$ NTs before and after thermal dewetting for various thicknesses of the Pt layer. Particularly, we found that in every case a maximized photocatalytic enhancement was obtained when sputtering-dewetting only 1-2 nm-thick Pt layers.[31] It can be clearly seen that if the layer was dewetted, an increase of the $H_2$ generation rate of the 1 nm Pt layer of more than 2 times was obtained compared with the as-sputtered Pt/$TiO_2$ NTs. The $H_2$ generation over time is linear (even in experiments carried out repeatedly), this indicating that these structures are stable (under UV illumination) even when Pt undergoes dewetting into few-nm large nanoparticles - in other words, the Pt clusters are well adherent to the photocatalyst and no co-catalyst flake off or deterioration takes place.

In general we found that regardless of the $TiO_2$ structure, the conversion of conformally sputtered films into Pt nanoparticles upon thermal dewetting in Argon was always observed (Figure S5, Supporting Information), and a clear benefit from thermal dewetting in terms of increase of $H_2$ evolution efficiency was obtained not only for the nanotubular oxide films but also for compact $TiO_2$ layers (Figure 1g).

In comparison with previous works on Pt/$TiO_2$ systems where fine grained Pt was uniformly deposited onto entire $TiO_2$ NTs (Figure S6 and 7, Supporting Information) by the classic photo-deposition method,[24,25] the present work shows how a minimal amount of co-catalyst (i.e., 1 nm-thick Pt film) can by an optimized sputtering-dewetting approach be structured to lead to a remarkable enhancement of the photocatalytic $H_2$ evolution rate.

Figure S8 (Supporting Information) shows the rate of photocatalytic $H_2$ evolution of Pt/$TiO_2$ NT arrays fabricated by photo-deposition, compared to the $H_2$ production of tubes decorated by our sputtering-dewetting approach.

Upon optimization of the photo-deposition method (see the Supporting Information for further discussion), we attained a *maximum* $H_2$ production of *ca*. 1.3 mL/5 h, which is slightly smaller than that achieved by sputtering-dewetting Pt atop the tubes (*ca*. 1.5 mL/5 h). Nevertheless, the key point here is the amount of Pt used for achieving this effect. Energy-dispersive X-ray spectroscopy (EDX) and inductively coupled plasma (ICP) measurements (Supporting Information) show that one of the best performing photo-deposited photocatalyst (i.e., "photo20-45min" in the Supporting Information) contains an amount of Pt of *ca*. 92 μg per sample, while the sputtering-dewetting of 1 nm-thick Pt layer led to tubes with a loading of Pt of *ca*. 2.1 μg, which is a nearly two-orders of magnitude lower loading of noble metal. In the light of these data, the effectiveness of the Pt deposition can be thus estimated by normalizing the $H_2$ evolution over the amount of loaded co-catalyst, as shown in Figure S8 (Supporting Information): this clearly highlights that our sputter-dewetting approach leads to a maximized photocatalytic performance by using minimal amount of Pt. Please, also note that even lower $H_2$ evolution rates were measured for the photo-deposited samples when the Pt amount was either lowered or increased (Figure S8, Supporting Information).

In general, the enhancement of $H_2$ generation by Pt decoration is ascribed to the formation of Schottky junction between Pt nanoparticles and $TiO_2$ and, as widely reported in the literature, this typically leads to a more efficient electron-hole separation.[18,20] The dewetting of 1-2 nm-thick Pt films on $TiO_2$ surfaces was shown to be optimal for $H_2$ evolution since it led to the formation of 4-10 nm-sized Pt nanoparticles and, at the same time, to an opening up of the underlying oxide to the environment and to incident light. Larger amounts of Pt (i.e., 2 to 10 nm-thick films) form much larger Pt particles (Figure S3, Supporting Information), and therefore lead to a larger shading effect of $TiO_2$, that is, to a covering of the oxide surface, which hinders the light absorption and consequently slows the photocatalytic process.[14]

However, an optimized sputtering-dewetting approach for 1 nm-thick Pt films on the NTs resulted in the increase of $H_2$ generation by *ca.* 250 times, compared to an improvement of only a factor ~ 5 observed for the compact $TiO_2$ layers. This is ascribed to different facts: *i)* the NTs clearly have larger surface area compared to compact layers, and this corresponds as well to a larger number of reaction sites; *ii)* the Pt nanoparticles on NTs are significantly smaller than those on compact layers, and this leads to a minimized shading effect along with an optimized density of the noble metal particles,[32] and *iii)* the expected orthogonal charge-separation and one-dimensional charge transport in the NTs may represent an additional key: compared to flat films, relatively thick layers of tubes (i.e., 6-7 µm) allow for maximized (full) light absorption and their enhanced charge-transport properties grant at the same time efficient electron migration over some µm-long distances towards the Pt clusters.[33]

As demonstrated above, Pt loadings and dewetting step are critical for achieving an enhanced photocatalytic $H_2$ generation. Furthermore, we found that the temperature of thermal dewetting is highly relevant when targeting efficient $H_2$ production, this because the structure and the phase composition of the $TiO_2$ tubes, as well as the dewetting process, are largely affected.

**Figure 2**a shows the X-ray diffraction (XRD) patterns of NT layers and compact oxides (see experimental section for the fabrication of the latter), with/without Pt decoration, all dewetted as outlined above. In every case, the samples show an intense peak at ~ 25°, corresponding to the formation of anatase. These results are in line with HR-TEM data shown in Figure S2

(Supporting Information) and discussed above. The effective Pt deposition on the NTs was also evident from the XRD pattern of the 10 nm Pt-modified oxide film that shows clear reflections at 46.3° and 67.6°, which correspond to the crystallographic parameters of Pt(0).[34]

Furthermore, X-ray photoelectron spectroscopy (XPS) characterization was carried out (Figure 2b), and Pt4f signals were observed for 1 nm Pt-decorated NTs and compact oxides at 71.58, 314.65 and 331.46 eV, that match the binding energy of Pt(0).[14,35,36] Here it is important to note that, as anticipated, besides inducing dewetting into monodispersed nanoparticles, the thermal treatment in Argon-atmosphere granted the sputtered Pt to retain its metallic state (i.e., Pt(0)),[27] both these features being essential for efficient conduction band electron trapping.[28]

**Figure 3**a shows the amounts of $H_2$ (5 h-long runs) generated with $TiO_2$ NTs that were loaded with 1 nm-thick Pt layers and thermally dewetted at different temperatures. The highest photocatalytic activity was measured for dewetting at 550 °C (Ar, 30 min), and corresponded to $H_2$ generation of nearly 1.5 mL.

Instead, annealing at 400 °C was found not to lead to Pt dewetting and the $H_2$ evolution was consequently lower. In fact, mainly depending on the Pt film thickness (and also on the annealing atmosphere), dewetting of Pt is expected to take place at around 550-600 °C, and therefore a temperature of *ca.* 400 °C is not sufficient for the process to occur.[37,38] Well in line with this, we found that the thermal treatment at T ≥ 600 °C could induce Pt dewetting but led also to extended sintering of the NT films which thereby also weakened their mechanical stability (Figure S9, Supporting Information). XRD patterns confirm that annealing at such high temperature led to the formation of a larger amount of rutile compared to the dewetting at 550 °C (Figure 3b). In line with previous reports,[39] this sintering phenomenon takes place by the growth of thermal oxide beneath the tube films, and by conversion of anatase into rutile $TiO_2$ (Figure 3c and d);[40–42] in other words, the loss of porosity and the formation of poorly photo-active rutile[43,44] are more likely the main cause for the dramatically lower $H_2$ evolution performance.

In summary, we illustrate a straightforward method based on a sputtering-dewetting strategy for decorating oxide films with metal co-catalyst nanoparticles. In view of efficient photocatalysis, the amount, chemical state, size and distribution of the Pt nanoparticles were shown to be of great importance, and could be properly controlled by tuning the sputtering-dewetting parameters. This approach, that can be easily extended to the deposition of other co-catalytic materials, led in the case of this study, to Pt/$TiO_2$ NT films with outstanding $H_2$ evolution (of more than 0.3 mL h$^{-1}$ cm$^{-2}$), particularly when only minimal amounts of sputtered Pt were used (only ca. 2.1 µg cm$^{-2}$). This is nearly 100 times lower loading - for the same $H_2$ evolution rate – than classic Pt photo-deposition approaches.

**Experimental Section**

*Growth of $TiO_2$ nanotubes*: Titanium foils (Advent Research Materials, 0.125 mm thickness, 99.6+% purity) were degreased by sonication in acetone, ethanol and deionized water, followed by drying in $N_2$ gas stream. The $TiO_2$ nanotubes were formed by anodizing titanium foils in ethylene glycol electrolyte containing $NH_4F$ (0.15 M) and $H_2O$ (3 wt%), at 60 V for 15 min (this led to tube layers with average thickness of 6-7 µm). The DC potential was applied by using a power supply VLP 2403 pro, Voltcraft. Right after the anodization, the samples were soaked in ethanol, and then dried under $N_2$ gas stream. Subsequently, the $TiO_2$ nanotubes were annealed at 450 °C in air for 1 h using a Rapid Thermal Annealer (Jipelec Jetfirst 100 RTA), with a heating and cooling rate of 30 °C min$^{-1}$. For preparing the $TiO_2$ compact oxides, titanium foils were anodized in $H_2SO_4$ (1 M) aqueous solution at 20 V for 15 min.

*Nanoparticle decoration*: The deposition of Pt on the anodic TiO$_2$ films was carried out using a high vacuum sputter coater (Leica - EM SCD500). The pressure of the sputtering chamber was reduced to $10^{-4}$ mbar, and then set at $10^{-2}$ mbar of Ar. The applied current was 17 mA. The amount of sputtered Pt was determined by an automated quartz crystal film-thickness monitor. Subsequently, the Pt decorated TiO$_2$ samples were annealed in Ar atmosphere at different temperatures. Dewetting could be induced when annealing at temperature higher than 400 °C. For the photo-deposition, chloroplatinic acid (H$_2$PtCl$_6$) was used as a metallic precursor. H$_2$PtCl$_6$ was dissolved in deionized water (0.9 mM) and then mixed with a water/ethanol mixture (65/35) with different volume ratios (2:98, 10:90, 20:80 and 40:60). The resulting platinum precursor solution was preliminary irradiated with UV light (medium pressure Hg lamp, 300 W) for 45 min in the absence of TiO$_2$. Subsequently, TiO$_2$ samples were placed in this solution in the dark for 3 h, followed by an irradiation step (UV light, for 45 or 2 min). The samples were then washed with deionized water and heat-treated in air at 120 °C for 2 h.

*Characterization of the structure*: Field-emission scanning electron microscope (FE-SEM, Hitachi S4800) and a high resolution transmission electron microscope (TEM, Philips CM300 UltraTWIN) equipped with a LaB6 filament and operated at 300 kV were used to characterize the morphology of the samples as well as to determine lattice spacing of anatase and rutile TiO$_2$. Energy-dispersive X-ray spectroscopy (EDAX Genesis, fitted to SEM chamber) was used for the chemical analysis. The chemical composition of the samples was analyzed by X-ray photoelectron spectroscopy (XPS, PHI 5600, US). X-ray diffraction (XRD) performed with a X′pert Philips MPD (equipped with a Panalytical X'celerator detector) was employed to examine the crystallographic properties of the materials.

*ICP measurements*: ICP-OES was carried out with a Spectro Ciros CCD. The samples were prepared by carefully dissolving the support/oxide films in a mixture of concentrated

HF/HNO$_3$/HCl 1/2/2 in volume (CAUTION: all the manipulation involving concentrated acids, especially HF, must be carried out wearing protective clothes and in a very efficient fume hood). The amount of Pt was detected using its emission line at 214.423 nm. As the amount to determine was generally low, the method of standard addition was followed. This avoided any matrix effect that could negatively affect the reliability of the determination. Two approaches were followed: firstly it was determined that the amount of deposited Pt was indeed constant for each specimen. To this purpose, 5 Pt loaded samples (ca. 128 mg each) were separately placed in 10 mL calibrated flasks and dissolved in 1 mL of acid, followed by diluting to the final volume with bi-distilled water. To 4 of the solutions, before the final dilution, an incremental addition of Pt standard (TRACECERT® 1000 mg L$^{-1}$) was done in order to get a final concentration of 0.1, 0.2, 0.3 and 0.4 mg L$^{-1}$, respectively. Reporting the intensity of the emission line at 214.423 nm vs the concentration increment resulted in a straight line ($R^2$ = 0.9828), confirming the amount of Pt loaded on each sample was constant, within the accuracy of the measurements. Then, on the basis of this evidence, the average amount of Pt loaded on the oxide/support was finally determined dissolving 5 samples in 5 mL of acid in a 25 mL calibrated flask. 4 aliquots of 5 mL of this solution were taken and diluted to 10 mL, and to 3 of them the Pt standard was incrementally added in order to get respectively a 0.5, 1.0 and 1.5 mg L$^{-1}$ final concentration. As blank for the background correction, we used a solution prepared dissolving 1 unloaded oxide/support film in 1 mL of acid and then diluting to 10 mL in a calibrated flask.

*Photocatalytic measurements*: Photocatalytic measurements were carried out by irradiating the oxide films with UV light (HeCd laser, Kimmon, Japan; λ = 325 nm, beam size = 0.785 cm$^2$, nominal power of 60 mW cm$^{-2}$) in a 20 vol% ethanol-water solution for 5 h in a quartz tube. The amount of produced H$_2$ (which accumulated over time within the tube) was measured by using a gas chromatograph (GCMS-QO2010SE, Shimadzu) equipped with a thermal conductivity detector and a Restek micropacked Shin Carbon ST column (2 m x 0.53 mm). GC measurements were carried out at a temperature of the oven of 45 °C (isothermal conditions), with the temperature of the injector set at 280 °C and that of the TCD fixed at 260 °C. The flow rate of the carrier gas, i.e., Argon, was 14.3 mL min$^{-1}$. The amount of evolved $H_2$ was measured always at the end of the experiments. Also for few checking runs, gas samples were withdrawn after 1, 3, 5 and 9 h to verify that the $H_2$ evolution was steady over time. Before the photocatalytic experiments, the reactor and the water-ethanol mixtures were purged with $N_2$ for 30 min to remove $O_2$. This is strictly needed as $O_2$ would diminish the efficiency of $H_2$ generation by competitively undergoing photocatalytic reduction to $O_2^{-\bullet}$ (that is, $O_2$ would react with conduction band electrons instead of water).

The photocatalytic experiments were carried out in ethanol-water solution since the presence of specific amounts of organics (i.e., methanol and ethanol) was proved to significantly trigger the $H_2$ generation. Exactly, ethanol acts as a hole-scavenger, meaning that the organic molecules are quickly oxidized towards several intermediates and finally to $CO_2$. As a result of the fast hole-consumption, conduction band electrons are more readily available for water reduction, thus yielding a larger amount of produced $H_2$.

**Acknowledgements**
The authors would like to acknowledge the ERC, the DFG and the DFG "Engineering of Advanced Materials" cluster of excellence for financial support as well as H. Hildebrand for valuable technical help. X. Zhou is acknowledged for XPS measurements and S. Özkan is acknowledged for the introduction of Pt photo-deposition.

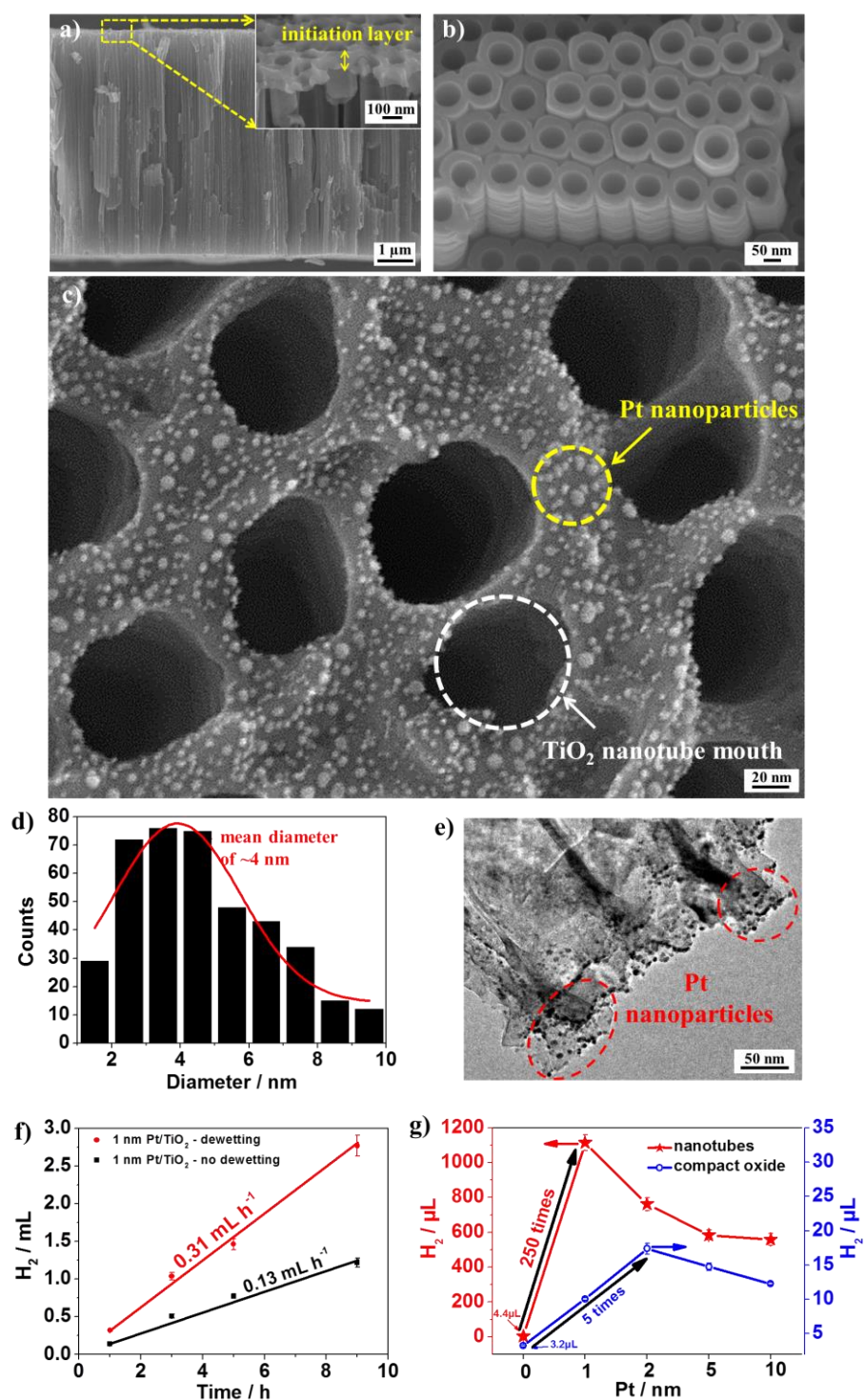

**Figure 1.** SEM images of: (a) and (b) as-formed $TiO_2$ NTs used in this work (the inset shows the initiation layer of the $TiO_2$ NTs); (c) $TiO_2$ NTs decorated with a 1 nm-thick Pt film and thermally dewetted at 550 °C in Ar. (d) Pt particle size distribution after thermal dewetting of 1 nm-thick Pt layer sputtered on $TiO_2$ NTs. (e) TEM image of $TiO_2$ NTs decorated with a 1 nm-thick Pt film thermally dewetted at 550 °C in Ar. Photocatalytic $H_2$ evolution measured for: (f) $TiO_2$ NT films decorated with 1 nm-thick Pt layers, before and after thermal dewetting at 550 °C in Ar; and (g) $TiO_2$ NT films and $TiO_2$ compact oxides loaded with different amounts of Pt and thermally dewetted in Ar at 450 °C for 30 min (all the photocatalytic runs lasted 5 h).

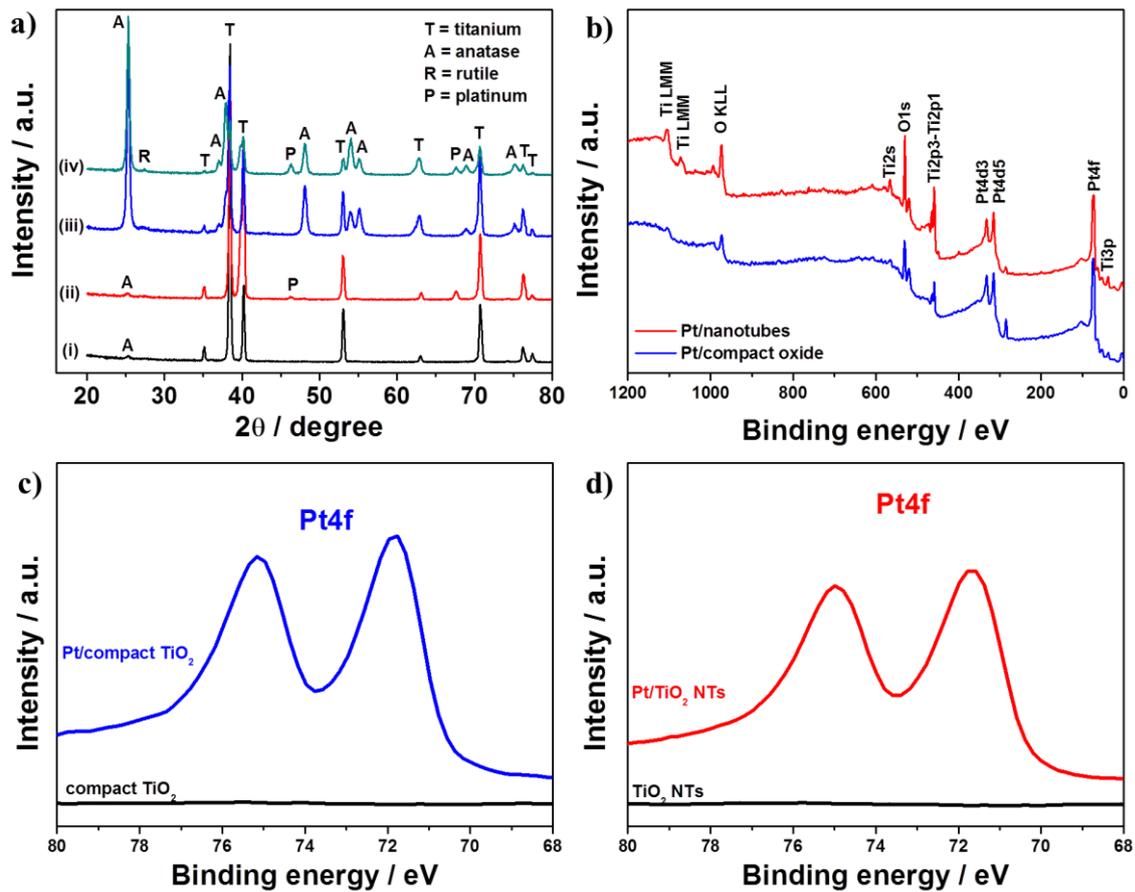

**Figure 2.** (a) XRD patterns of (i) compact oxide, (ii) 10 nm-Pt/compact oxide, (iii) $TiO_2$ nanotubes and (iv) 10 nm-Pt/nanotubes. (b) XPS survey of 1 nm-Pt/compact oxide and 1 nm-Pt/nanotubes. XPS spectra in the Pt4f region of: (c) compact oxide and 1 nm-Pt/compact oxide; (d) $TiO_2$ nanotubes and 1 nm-Pt/nanotubes.

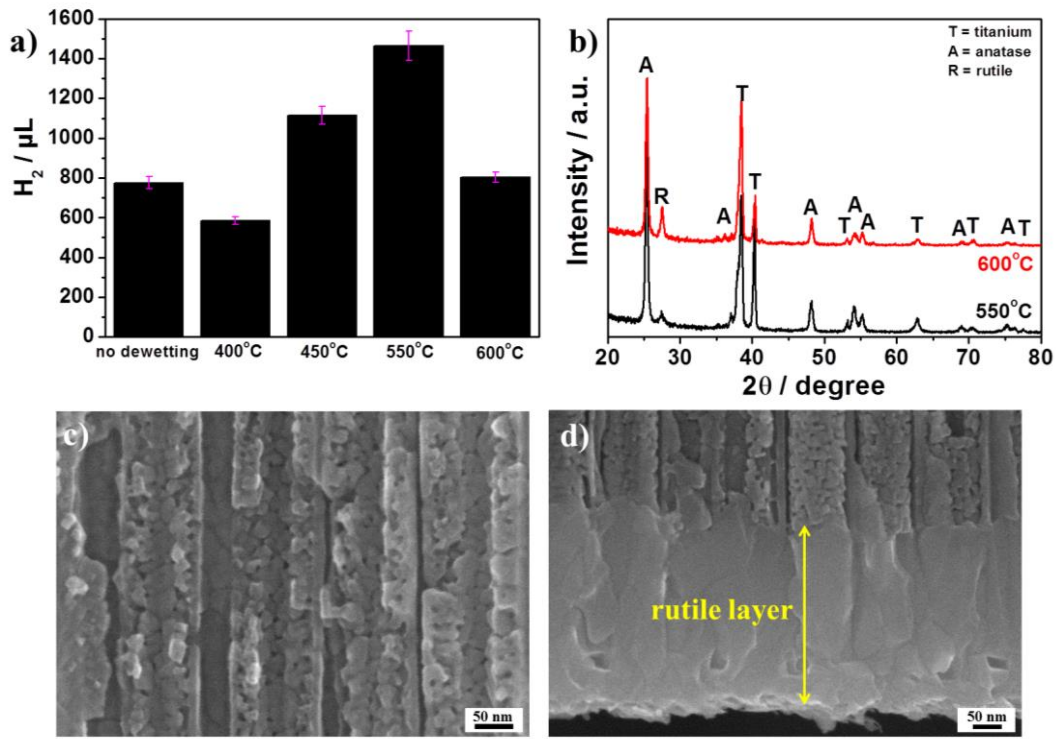

**Figure 3.** (a) Photocatalytic H$_2$ evolution measured for TiO$_2$ NTs loaded with 1 nm-thick Pt layers and thermally dewetted in Ar at different temperatures. (b) XRD patterns of TiO$_2$ NT films decorated with 1 nm-thick Pt layers after thermal dewetting at 550 and 600 °C in Ar. SEM images of TiO$_2$ NT films decorated with 1 nm-thick Pt layers after thermal dewetting at 600 °C in Ar: (c) cross-section and (d) rutile layer at the bottom of NTs.

**The least is the best**. We introduce a technique to strongly reduce Pt use for photocatalytic hydrogen generation from $TiO_2$ nanotubes. By site-selectively depositing thin layers of Pt only at the "mouth" of the nanotubes and by a subsequent thermal dewetting step, we achieve an outstanding photocatalytic improvement with minimal amounts of co-catalyst.

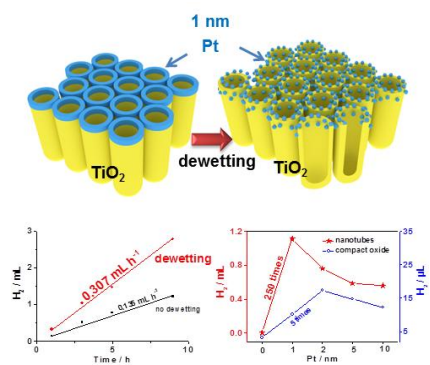

# Supporting Information

**Noble metals on anodic TiO$_2$ nanotubes mouths: Thermal dewetting of minimal Pt co-catalyst loading leads to significantly enhanced photocatalytic H$_2$ generation**

*Nhat Truong Nguyen, Marco Altomare, JeongEun Yoo, Nicola Taccardi, and Patrik Schmuki\**

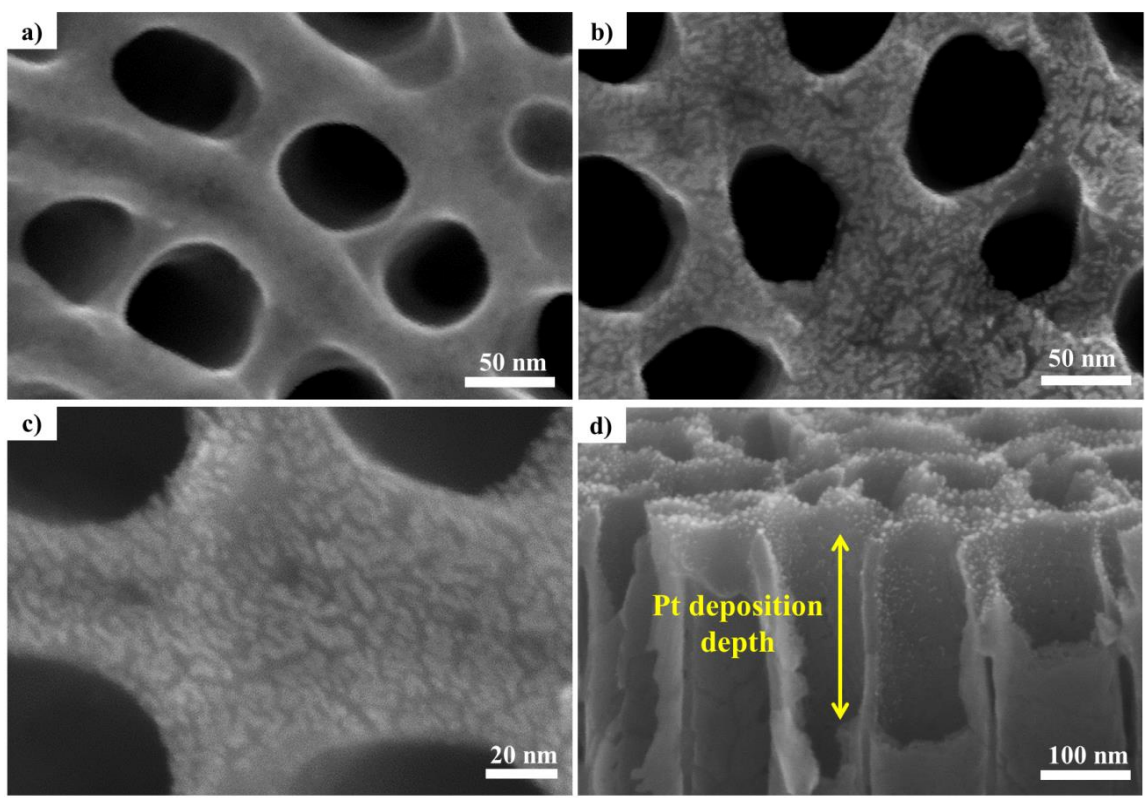

**Figure S1.** SEM images of: (a) bare $TiO_2$ NTs; $TiO_2$ NTs decorated with a 1 nm-thick Pt film (b), (c) before and (d) after thermal dewetting.

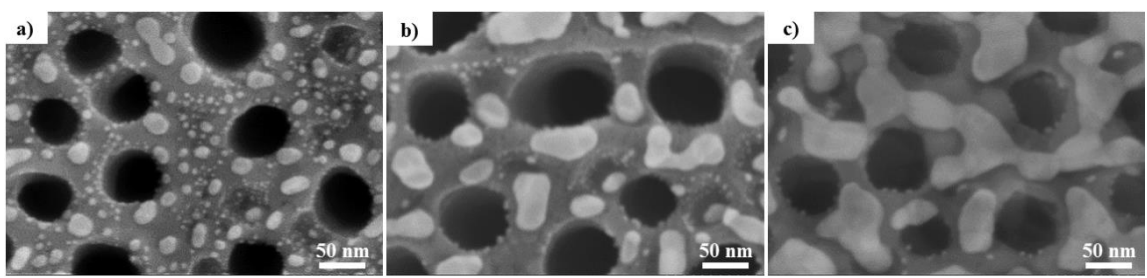

**Figure S2** SEM images of $TiO_2$ NTs decorated with (a) 2, (b) 5 and (c) 10 nm-thick Pt films after thermal dewetting.

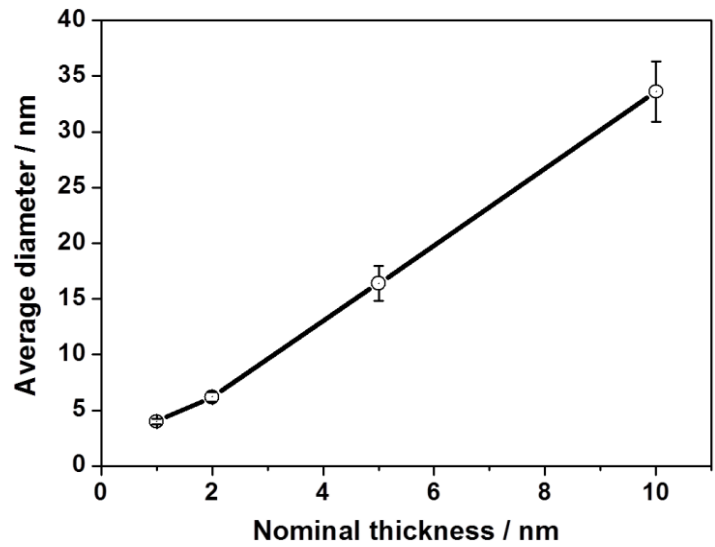

**Figure S3** Dependence of the nanoparticle size on the nominal thickness of the sputtered Pt layers.

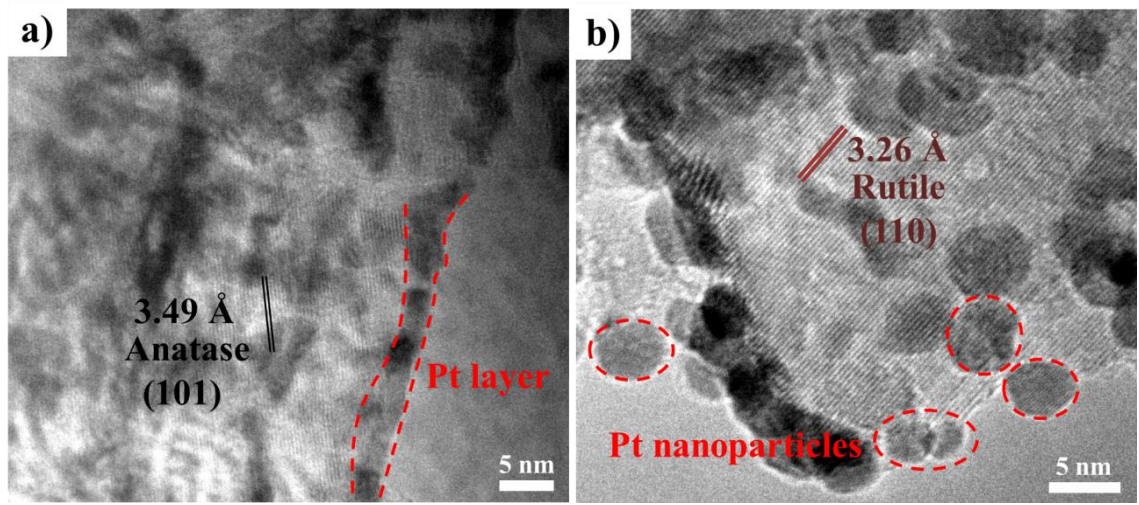

**Figure S4** HR-TEM images of TiO$_2$ NTs loaded with 1 nm-thick Pt layers (a) before and (b) after thermal dewetting in Ar at 550 °C.

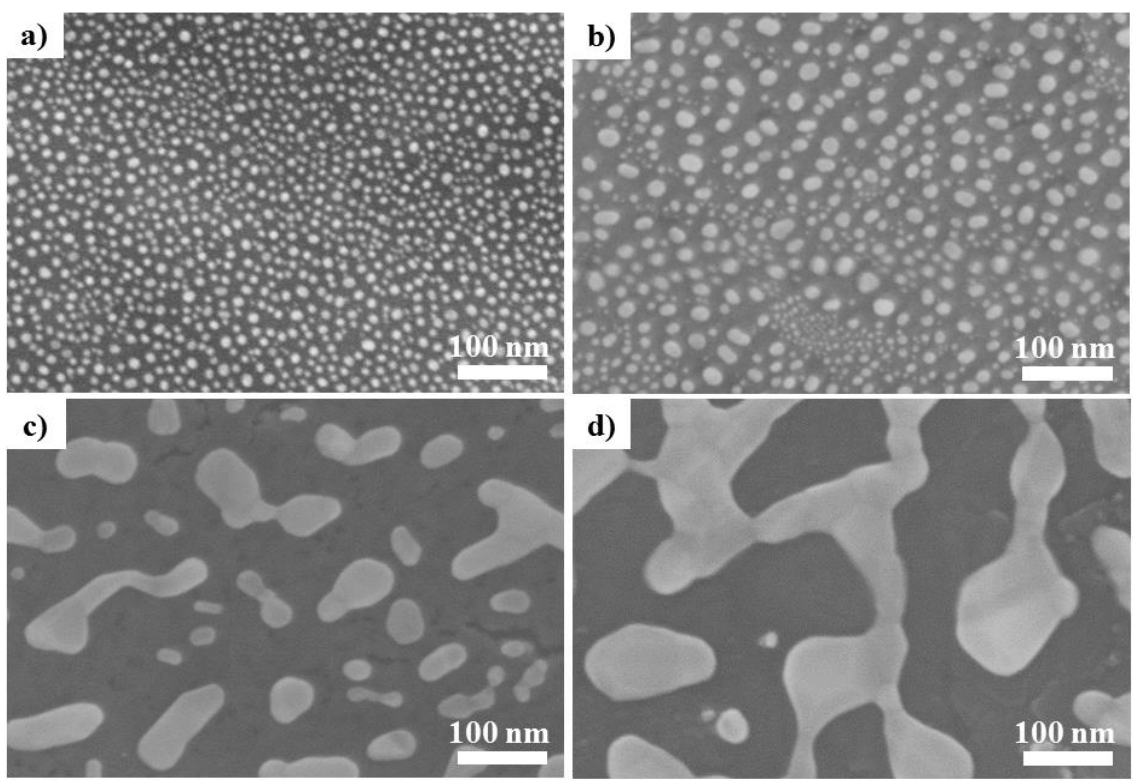

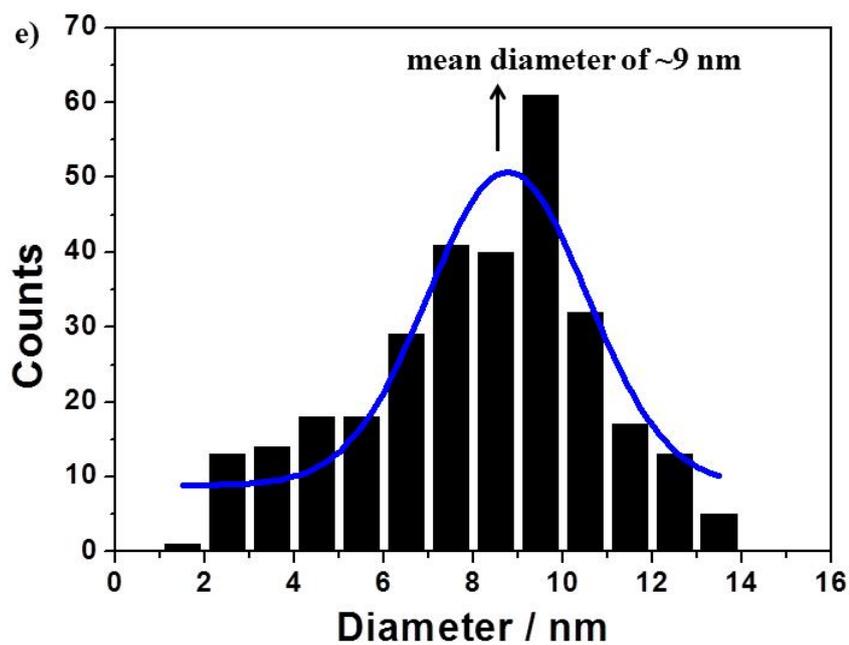

**Figure S5** SEM images of TiO$_2$ compact layers loaded with (a) 1, (b) 2, (c) 5 and (d) 10 nm-thick Pt layers thermally dewetted in Ar at 450 °C. (e) Pt nanoparticle size distribution after thermal dewetting of 1 nm-thick Pt layers sputtered on a compact oxide film.

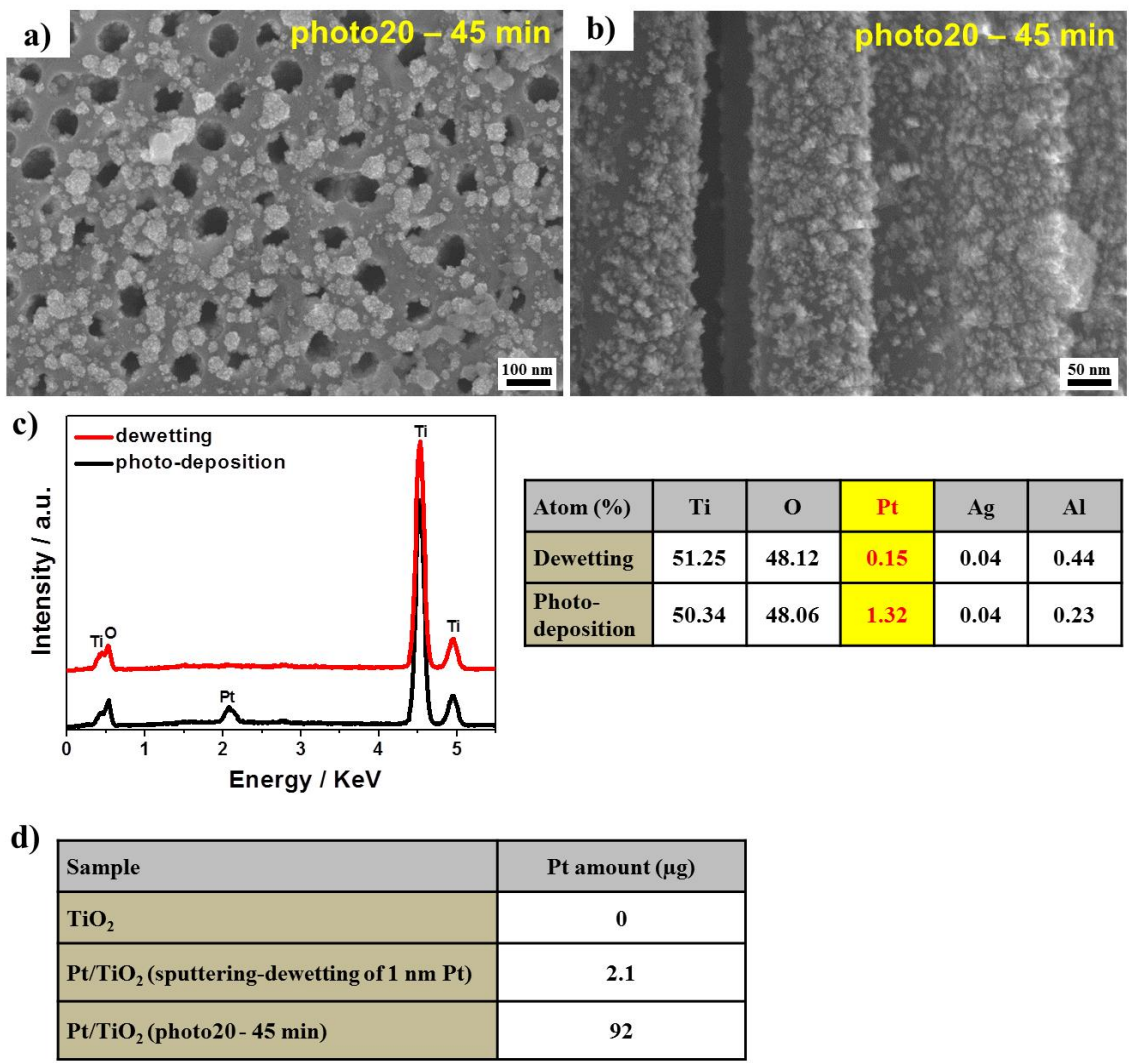

**Figure S6** (a) and (b) SEM images of $TiO_2$ nanotubes loaded with Pt by photo-deposition method. (c) EDX and (d) ICP measurements of $TiO_2$ nanotubes loaded with Pt by sputtering-dewetting and photo-deposition methods. The presence of aluminum detected by EDX is ascribed to the aluminum sample holder and the peak of Ag is ascribed to silver paste.

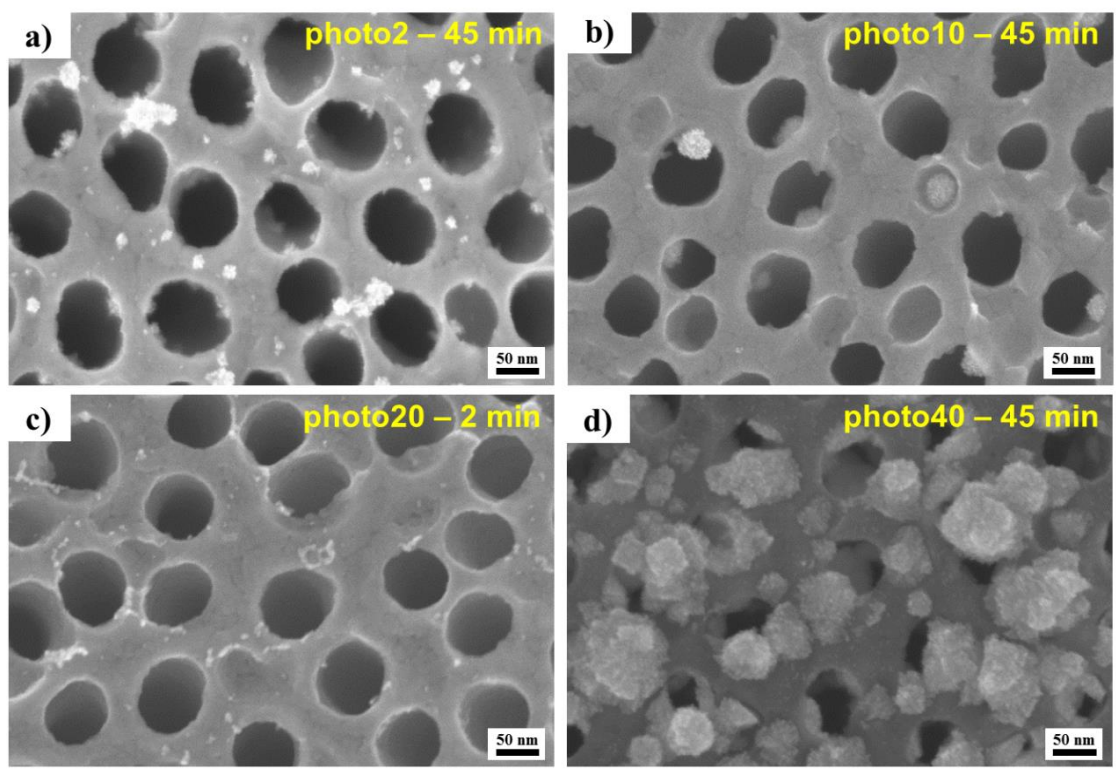

**Figure S7** SEM images of TiO$_2$ nanotubes loaded with different Pt amounts by photo-deposition method: (a) 2, (b) 10, (c) 20 and (d) 40 vol%. The samples were irradiated with UV light for 45 min, except for sample c) that was irradiated for 2 min.

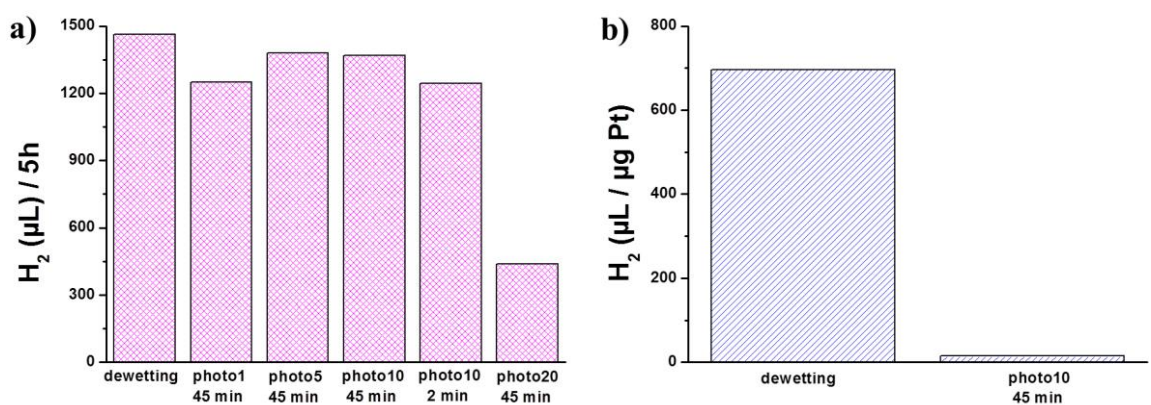

**Figure S8** Photocatalytic H$_2$ evolution measured for TiO$_2$ nanotubes loaded with Pt nanoparticles by sputtering-dewetting and photo-deposition methods (UV light irradiation time is 5 h for all the photocatalytic runs): (a) shows the absolute amount of evolved H$_2$, while (b) shows the amount of H$_2$ normalized over the co-catalyst (Pt) amount.

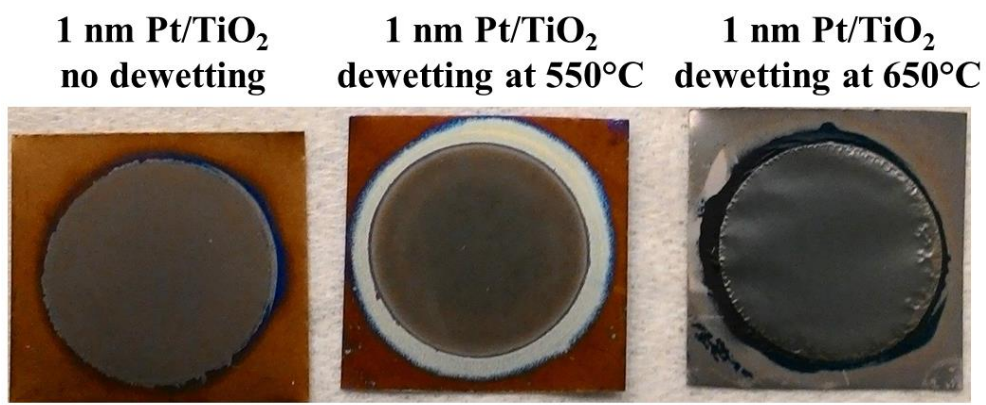

**Figure S9** Optical images of $TiO_2$ NT layers loaded with 1 nm-thick Pt films before and after thermal dewetting (in Ar at 550 and 650 °C).